\documentclass[a4paper]{jpconf}
\usepackage{graphicx}
\usepackage{amsmath,amssymb,amsfonts}
\usepackage{subfigure}
\usepackage{listings}
\usepackage{cite}
\def\gv{\mbox{GeV}}
\def\tv{\mbox{TeV}}
\def\msbar{$\overline{\mbox{MS}}$}

\begin{document}
\title{Self-consistence of the Standard Model via the renormalization group analysis}

\author{Fred~Jegerlehner$^{1,2}$,  Mikhail~Kalmykov$^{3,4}$ and Bernd~A.~Kniehl$^3$} 
\address{$^1$ 
Humboldt-Universit\"at zu Berlin, Institut f\"ur Physik,
Newtonstra\ss e 15, 12489 Berlin, Germany}
\address{$^2$ 
Deutsches Elektronen-Synchrotron (DESY),
Platanenallee 6, 15738 Zeuthen, Germany}
\address{$^3$ II. Institut f\"ur Theoretische Physik, Universit\"at Hamburg,
Luruper Chaussee 149, 22761 Hamburg, Germany}
\address{$^4$ Joint Institute for Nuclear Research,
$141980$ Dubna (Moscow Region), Russia} 

\ead{mikhail.kalmykov@desy.de}

\begin{abstract}
A short review of recent renormalization group analyses of the 
self-consistence of the Standard Model is presented.
\end{abstract}
\vspace{-1cm}
\section{Introduction}
The recent discovery of the Higgs boson \cite{ATLAS,CMS} at the LHC with mass 
\begin{eqnarray}
M_H & = & 125.03^{+0.26}_{-0.27} \mbox{(stat)} {}^{+0.13}_{-0.15} \mbox{(syst)}~\gv \;,
\quad 
\mbox{\cite{CMS:2014}}
\nonumber \\ [0.3cm]
M_H & = & 125.36 \pm 0.37 \mbox{(stat)} \pm 0.18 \mbox{(syst)}~\gv \;,
\quad 
\mbox{\cite{ATLAS:2014}} \;, 
\nonumber 
\end{eqnarray}
and the fact that so far no signal for new physics has been found,
gives rise to the necessity to analyze in detail the self-consistency
of the Standard Model.  The Higgs boson is a necessary ingredient of
the Standard Model (SM), required by its perturbative
renormalizability \cite{tHooft:1971,HV1972}.  However, the
renormalizability does not give rise to any constraints on the values
of parameters of the Lagrangian, but only on the type of interactions.
Some additional restrictions on the values of masses and coupling
constants are coming from {\it unitarity}, {\it triviality} and {\it
vacuum stability}.  These three ideas have been used within last
decades for an analysis of the self-consistency of the SM.


\vspace{0.2cm}
\noindent
{\it Unitarity bound}:
bounds on masses of fermions and Higgs bosons of any renormalizable model 
can be derived from considerations of the radiative corrections to decays 
and/or scattering processes \cite{Dicus:1973,Lee:1977}.
The breakdown of unitary can be avoided by adding a new particles
(which implies the existence of new physics) or by the requirement
that the perturbative approach must be meaningful 
(this implies a bound on the particles mass or its coupling constants) \cite{Veltman:1977}.
Unitary bounds depend on the type of process under consideration and 
the precise definition of the breakdown of the perturbative approach.
For the Standard Model, the unitary bound for the Higgs boson mass is
$
M_H \lesssim 1~\tv \;.
$

\noindent
{\it Triviality and Landau pole}: 
The {\it triviality} constraint is related with the high energy behavior of 
the running couplings.
Is it well known, that a running coupling $h(\Lambda)$ may suffers from 
a Landau pole when the corresponding $\beta$-function is positive: 
\begin{equation}
\frac{d }{d \Lambda} h(\Lambda) = \beta_0 h^3(\Lambda)
 \longrightarrow 
h^2(\Lambda) = \frac{h_R^2}{1 - \beta_0 h_R^2 \ln \frac{\Lambda}{M}} \;,
\label{Landau}
\end{equation}
where $\Lambda$  is the ultra-violet cutoff, 
$M$ is a characteristics scale of the process under consideration
and $h_R$ is the renormalized (running) charge at scale $M$.
As follows from Eq.~(\ref{Landau}), 
the running coupling $h^2(\Lambda)$ diverges at scale $\Lambda_{\mbox{Landau}}$  
defined as
$
\Lambda_{\mbox{Landau}}  =M\,\exp \left[ \frac{1}{\beta_0 h_R^2}\right] \;, 
$  
when $\beta_0 > 0$.
In the SM, Landau poles may exists for the 
$U(1)$-gauge coupling, $g_1$,
Yukawa couplings, $y_f$ and the Higgs self-coupling, $\lambda$. For
the most important couplings the SM renormalization group (RG) equations read 
$$
\frac{d h}{dt} = \frac{1}{16 \pi^2} \beta_h  \;, \quad h \in \{g_1, g_2, g_3, y_f, \lambda \} \;, 
$$
where in the one-loop approximation the corresponding $\beta$-functions have the following form: 
\begin{eqnarray}
&& 
\beta_1 =    \frac{41}{6} g_1^3 \;, \quad 
\beta_2 =  - \frac{19}{6} g_2^3 \;, \quad 
\beta_3 = - 7 g_3^3 \;, 
\quad 
\beta_{y_t} = y_t
\biggl[ 
\frac{9}{2} y_t^2 
- 8  g_3^2 
- \frac{9}{4} g_2^2 
- \frac{17}{12} g_1^2 
\biggr] \;,   
\nonumber \\ && 
\beta_\lambda  =   
\lambda
\biggl[
24 \lambda 
\!+\! 12 y_t^2
\!-\! 9 g_2^2 
\!-\! 3 g_1^2 
\biggr] 
-
{ 
\biggl[
        6 y_t^4
\!-\!  \frac{9}{8} g_2^4 
\!-\! \frac{3}{8} g_1^4
\!-\! \frac{3}{4} g_1^2 g_2^2 
\biggr] 
}
\;. 
\label{SM:coupling}
\end{eqnarray}
It has been shown in \cite{Riesselmann:1997} that the Landau pole in the 
Higgs self-coupling $\lambda$ would be below Planck scale 
if the Higgs is heavier than $180~\gv$.
The results of recent analyses \cite{Jegerlehner:2014,buttazzo2013}
have confirmed that all coupling constants are free from 
Landau singularities and have smooth behavior
in interval between $M_Z \sim 90~\gv$ and $M_{Planck}=2.435 \times 10^{18}~\gv$ 
(see Fig.~\ref{running}):
\begin{figure}[h]
\centerline{ 
  \includegraphics[width=4.45cm]{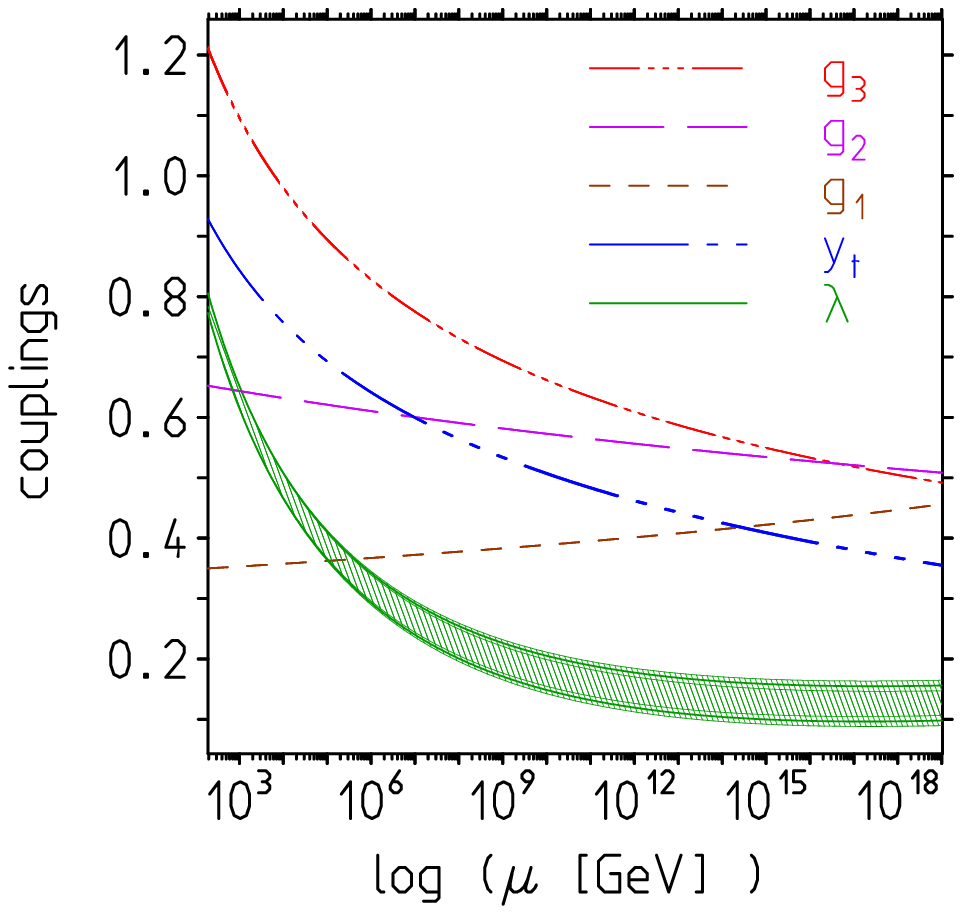}
  \hspace*{5mm}
  \includegraphics[width=4.45cm]{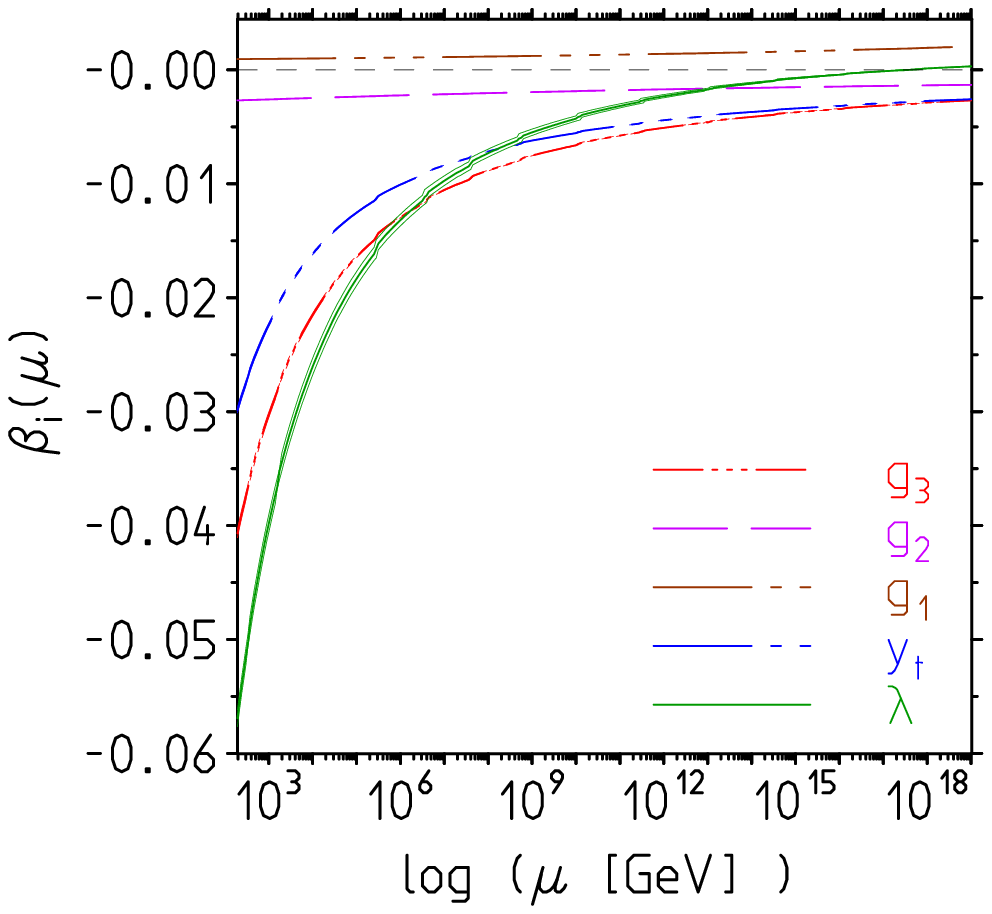}
            }
\vspace*{-3mm}
\caption{
The running coupling constants(left) and corresponding $\beta$-functions (right).
The plots are taken from \cite{Jegerlehner:2014}.
}
\label{running}
\end{figure}

\vspace{-0.5cm}
\subsection{Vacuum stability and effective potential}
For the analysis of the {\it vacuum stability} one needs to recall some basic definitions related with 
the scalar potential and its generalization in Quantum Field Theory.
Let us remind, that at the classical level, the Higgs potential in the SM  
\begin{equation}
V(\phi) = \frac{1}{2} m^2 \phi^2 + \frac{1}{4!} \lambda \phi^4 \;, 
\label{potential}
\end{equation}
is bounded from below for $\lambda > 0$
and has a trivial minimum for $m^2 > 0$ at $\phi=0$, and 
non-trivial minima at $\phi^2 = \frac{-m^2}{\lambda}$
for $m^2 < 0$.
At the quantum level, instead of the classical potential defined by Eq.~(\ref{potential}),  
the {\it effective potential} should be analysed \cite{Coleman:1973}.
It can be written in the Landau gauge and the \msbar~scheme as \cite{Ford:1993}
\begin{eqnarray}
V(\phi) & \to & V_{eff}(\phi(t)) = 
          - \frac{1}{2} m^2(t)\phi^2(t)
          + \frac{1}{4!} \lambda(t) \phi^4(t)  
          + \frac{1}{16 \pi^2} V_1 
          + \frac{1}{(16 \pi^2)^2} V_{rest} 
\;, 
\end{eqnarray}
where (we use the notations of \cite{Quiros:1995}): 
\begin{eqnarray}
V_1 & = &  \sum_i \frac{k_i}{4} M_i^4(\phi(t)) \left[\ln \frac{M_i^2(\phi(t))}{\mu^2}  - C_i \right] \;, 
\nonumber \\ 
\phi(t) & = & \exp \{ - \int_0^t \gamma(\tau) d\tau \} \phi_{clas}  \;, 
\quad 
M_i^2(\phi(t))  =  a_i h_i(t) \phi(t)^2 + b_i  \;,
\end{eqnarray}
$h_i(t)$ are the running couplings,  
$\gamma(t)$ is the anomalous dimension of the Higgs field,
and  
$k_i, C_i, a_i, b_i$ are numerical constants. 
By $V_{rest}$ we denote the higher-order contributions, 
which, in particular, also include the higher dimension operators 
(it begins at four loop) \cite{Ford:1993,Nakano:1993jq}:
\begin{eqnarray}
V_{rest} \sim \lambda \phi^4 
\sum_{L>4} \left( \frac{\lambda^2 \phi^2}{m^2+\frac{1}{2}\lambda \phi^2} \right)^{L-3} \;, 
\label{rest}
\end{eqnarray}
where $L$ is the number of loops (for more details see Section 4 of \cite{Ford:1993} or \cite{EJ:2007}).

\begin{figure}[h]
\centerline{\includegraphics[width=0.6\textwidth]{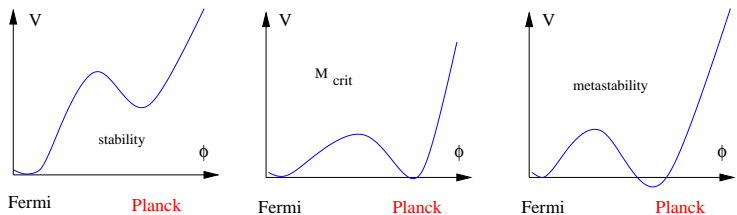}}
\caption{
The form of the effective potential for the Higgs field $\phi$ which
corresponds to the stable (left), critical (middle) and metastable
(right) electroweak vacuum. 
The plot is taken from \cite{Shaposhnikov:2013}.
}
\label{stab}
\end{figure}

The quantum corrections modify the shape of the effective potential
such that a second minimum at large (Planck) scale may be generated.
This second minimum (see Fig.~(\ref{stab})) is: 
(left plot)
stable  
$V_{eff}(v) < V_{eff}(\phi_{\mbox{min}})$, 
(middle plot)
critical (two minima are degenerate in energy) 
$V_{eff}(v) = V_{eff}(\phi_{\mbox{min}})$, 
(right plot) 
unstable/metastable 
$V_{eff}(v) > V_{eff}(\phi_{\mbox{min}})$, 
electroweak vacuum and 
%
%
\vspace{-0.5cm}
\noindent 
$v$ is the location of the EW minimum 
and $\phi_{\mbox{min}}$ is the value of a new minimum.
Depending on the values of the Higgs boson and top quark masses
the lifetime of the EW vacuum can be larger or smaller 
then the age of the Universe \cite{vacuum:1,vacuum:2,vacuum:3}.
The first case corresponds to the metastability scenario.

To get {\it practical} criteria of stability, 
the following step-by-step approximations are used \cite{Espinosa:2013}:  
(i) the potential at very high values of the field is dominated by
the quartic term 
$
V_{eff} \sim  \lambda(\phi) \phi^4 
$
and $\lambda(\phi)$ depends on $\phi$ as the running coupling $\lambda(\mu)$
depends on the running scale $\mu$; 
(ii) 
it's looking the large value of the field $\phi=\phi_{crit}$, where  $\lambda(\phi_{crit})=0$;
(iii) 
combine together the previous two approximations, 
we get at very high values of the Higgs field
$
\left.
V_{eff}
\right|_{\phi \sim \phi_{crit}} \simeq  \lambda(\phi_{crit}) \phi_{crit}^4.
$
Then, the effective potential becomes negative (unbounded from below) when $\lambda(\mu) < 0$ 
and the vacuum at the EW scale is not the absolute minimum.   
As follows from Eqs.~(\ref{SM:coupling}), 
the Higgs self-coupling $\lambda$ is the only SM dimensionless
coupling that can change sign with the scale variation 
since its beta-function, $\beta_\lambda$ contains a part which is not proportional to  $\lambda$. 
In the considered approximation,  \\
{\it  
the requirement that the electroweak vacuum 
$<\phi> = (\sqrt{2} G_F)^{-\frac{1}{2}} = 246.22~\gv$
is the absolute minimum of the potential 
up to scale $\Lambda$, implies 
\begin{equation}
\lambda(\mu)>0 \;, \quad \mbox{for any} \quad \mu<\Lambda \;.
\end{equation}
}
\\
The role of quadratic term or higher dimensional terms (see Eq.(\ref{rest}))
is discussed in Section 3.

\vspace{-0.5cm}
\section{Analysis}
\subsection{Renormalization group equations and Matching Conditions}
Both the questions, the one concerning vacuum stability and the one
concerning triviality, in the Standard Model 
are reduced to the renormalization group analysis of the Higgs self-coupling $\lambda$. 
The evolution of $\lambda$ includes the evolution of all coupling constants,
see Eq.~(\ref{SM:coupling}). 

The starting point for solving the evolution equations is provided by
the matching conditions: 
the relations between the running coupling constants $h_i(\mu)$
and the relevant (pseudo-) physical observables, $\sigma_i$.  
The simplest form for the matching conditions follows when physical masses are taken 
as the referring point: $\sigma_i \equiv M_i$. In this case the matching conditions 
have the following form \cite{Sirlin:1986}: 
\begin{eqnarray}
h_i(\mu^2) = c_i \frac{G_F}{\sqrt{2}} M_i^2 
\left( 1 + \delta_{\alpha_s} + \delta_\alpha + \delta_{\alpha \alpha_s} 
         + \delta_{\alpha^2} + \delta_{\alpha_s^2} + \cdots \right)  \;, 
\label{mass}
\end{eqnarray}
where 
$c_i$ are normalization constants and 
$\delta_{\alpha_j^k}$ are including only propagator type diagrams of order $O(\alpha_j^k)$.
The evaluation of the one-loop EW matching conditions of order $O(\alpha)$
was partially done in \cite{Sirlin:1986,Sirlin,FJ1981} 
and has been completed in \cite{BHS86,Yukawa:1}.
The two-loop matching conditions of order $O(\alpha^2)$ 
for the Higgs self-coupling $\lambda$ in the limit of heavy Higgs boson 
have been evaluated in \cite{vanderBij:1983,Ghinculov:1994}, 
and full two-loop results were completed in \cite{Awramik,Higgs1,Higgs2,Higgs3,Higgs3b,Higgs4}.
The $O(\alpha \alpha_s)$ corrections for the top-quark Yukawa coupling
within the SM
have been evaluated in \cite{JK2003,JK2004} and $O(\alpha^2)$ corrections 
are available only in the gaugeless limit \cite{fermion,KV2014}.
A state of the art evaluation attempting matching at the two-loop
level has been reported in ~\cite{buttazzo2013} (results see below).

The evaluation of the 3-loop RG equations for the SM gauge coupling
was started in \cite{KTV,TVZ}, and full 3-loop results for the
Standard Model have been published in
\cite{Mihaila:2012a,Mihaila:2012b,Chetyrkin-Zoller-2012,
      Bednyakov:2013a,Bednyakov:2013b, Bednyakov:2013c,
      Chetyrkin-Zoller-2013}.

\subsection{Results of RG analysis}
After the top-quark discovery
the results of RG analyses of self-consistence of the SM are typically 
fitted by three parameters and can be written as follows:
\begin{equation}
M_{\mbox{min}} > \left[M_{crit} \!+\! c_1 \times \left(M_t-M_{\mbox{exp}} \right) 
                     \!+\! c_2 \times \left(\alpha_s-\alpha_{s,\mbox{exp}}\right) \right]~\gv \;, 
\label{fit}
\end{equation}
where 
$M_{crit}$ is the critical value of the Higgs boson mass, 
$c_1, c_2$ are some numerical coefficients
and $M_{\mbox{exp}}$ and $\alpha_{s,\mbox{exp}}$ 
are the latest experimental values of the pole mass of the top-quark
and of the strong coupling constant, 
respectively.

The detailed analysis of stability/metastability bounds 
based on the two-loop RG equations for all SM couplings and one-loop EW matching conditions 
(as well as two-loop QCD corrections to the top-quark Yukawa coupling $y_t$)
have been presented in \cite{last-prediction}:
\begin{equation}
M_{\mbox{min}} \geq
\left[
130
\!+\! 1.8\! \times \! \left( \frac{M_t^{\rm pole} \!-\! 173.2~{\rm GeV}} {0.9~{\rm GeV}}
\right)
\!-\! 0.5 \! \times \! \left( \frac{ \alpha_s (M_Z) \!-\! 0.1184}{0.0007}  \right) 
\pm 3
\right]
~\gv \;, 
\label{2RG}
\end{equation}
with
$M_t=(173.2 \pm 0.9)~\gv$ and $\alpha_s(M_Z)=0.1184 \pm 0.0007$
as input parameters and where the error of $3~\gv$ is an estimation 
of unknown higher-order corrections.

The analysis of vacuum stability performed with inclusion of 3-loop RG equations
and 2-loop matching conditions was presented in \cite{Bezrukov:2012}
and \cite{Degrassi:2012} with the following results:

\begin{eqnarray}
M_{\mbox{min}} & \geq & 
\left[
128.95
\!+\! 2.2\! \times \! \left( \frac{M_t^{\rm pole} \!-\! 172.9~{\rm GeV}} {1.1~{\rm GeV}}
\right)
\!-\! 0.56 \! \times \! \left( \frac{ \alpha_s (M_Z) \!-\! 0.1184}{0.0007}  \right) 
\pm 1
\right]
~\gv \;, 
\quad \mbox{\cite{Bezrukov:2012}}
\nonumber \\ 
\label{bezrukov}
\\
M_{\mbox{min}} & \geq & 
\left[
129.4
+ 1.4\! \times \! \left( \frac{M_t^{\rm pole} \!-\! 173.1~{\rm GeV}} {0.7~{\rm GeV}}
\right)
- 0.5 \! \times \! \left( \frac{ \alpha_s (M_Z) \!-\! 0.1184}{0.0007}  \right) \pm 0.7 \right]
~\gv \;,
\quad \mbox{\cite{Degrassi:2012}}
\nonumber \\ 
\label{degrassi}
\end{eqnarray}

\noindent
{\it The latest published analysis} have been presented in \cite{buttazzo2013}:
\begin{eqnarray}
M_{\mbox{min}} > 
\left[ 129.6
\!+\! 2.0\! \times \! \left( M_t^{\rm pole} \!-\! 173.35~{\rm GeV} \right)
\!-\! 0.5 \! \times \! \left( \frac{ \alpha_s (M_Z) \!-\! 0.1184}{0.0007}  \right) 
\pm 0.3
\right]
~\gv \;.
\label{full}
\end{eqnarray}

\subsection{The analysis of uncertainties}
As follows from Eqs.~(\ref{2RG})-(\ref{full}) 
adding two-loop matching conditions and 3-loop RG equations 
does not changed dramatically the central value of critical mass: $130 \to \sim 129.0 \div 129.6$,
but essentially reduces the theoretical uncertainties: $(3 \to 0.3)~\gv$.
Adding the leading 3-loop corrections to the matching conditions for the Higgs self-coupling 
\cite{3-loop_1,3-loop_2} does not modified the final result, Eq.~(\ref{full}).
The $3\sigma$ variations of $\alpha_s$ or the small variation of the value of
the matching scale \cite{Masina:2012} 
also do not produce the significant corrections to the fitted expression.

{\it The main error} in Eqs.~(\ref{2RG})-(\ref{full}) is coming from 
the input value of top-quark mass:
$
\Delta M_{\mbox{min}} \sim  2.0\! \times \! \Delta M_t^{\rm pole}. 
$
Currently the most precise measurement of the top-quark mass 
has been reported as the world combination of the ATLAS, CDF, CMS and D0 \cite{top}
(for completeness we present also the result based only on CDF/D0 data \cite{top:2012}): 
\begin{eqnarray}
M_t  & = &  173.18 \pm 0.94~\gv \;,  \quad \mbox{\cite{top:2012}} 
\\
M_t  & = &  173.76 \pm 0.76~\gv \;,  \quad \mbox{\cite{top}}
\label{quark}
\end{eqnarray}
It has been pointed out in \cite{Hoang:2008}
that the values of the top quark mass quoted by the experimental collaborations 
correspond to parameters in Monte Carlo event generators in which, 
apart from parton showering, the partonic subprocesses are calculated at the tree level, 
so that a rigorous theoretical definition of the top quark mass is lacking 
(see more detailed discussion of this issue in \cite{Sbornik}).
In particular, the following issues in precision top mass determination at hadron colliders are relevant
\cite{Juste:2013}:
MC modeling, 
reconstruction of the top pair,
unstable top and finite top width effects,
bound-state effects in top pair production at hadron colliders,
renormalon ambiguity in top mass definition,
alternative top mass definitions,
higher-order corrections,
non-perturbative corrections,
contributions from physics beyond the Standard Model.

To reduce the uncertainties related with undetermined differences between 
Monte-Carlo and pole masses (it was estimated in \cite{Hoang:2008,Sbornik} as $1~\gv$)  
the mass of top-quark can be extracted directly
from a measurement of the total top-pair production cross section 
$\sigma_{\mbox{exp}}( p\bar p \to t\bar t\! + \!X)$. 
Such analysis performed in \cite{ADM} with NNLO accuracy
with inclusion of the full theoretical uncertainties
(the scale variation as well as the (combined) PDF and $\alpha_s$ uncertainties)  
gives rise to the following result, 
${ M_t^{\rm pole}  =  173.3 \pm 2.8~\gv} \;.$
The central value is very close to the one in Eq.~(\ref{quark}), 
but the theoretical uncertainty is much larger.
Similar analyses 
(direct extraction of the pole mass of top-quark from measured total cross section)
have been performed by a few other groups \cite{TOPIXS,TOPIXS2013,Top++,ABM:2013,Sbornik}
with the following results:
\begin{eqnarray}
M_t^{\rm pole}  & = & 171.4^{+5.4}_{-5.7} ~\gv \quad \mbox{\cite{TOPIXS}} \;, 
\label{m1}
\\ 
M_t^{\rm pole}  & = & 174.3^{+4.9}_{-4.4} ~\gv \quad \mbox{\cite{TOPIXS2013}}  \;, 
\\ 
M_t^{\rm pole}  & = & 174.2^{+3.6}_{-3.9} ~\gv \quad \mbox{\cite{Sbornik}}  \;, 
\\
M_t^{\rm pole}  & = & 171.2 \pm 2.4 \pm 0.7~\gv \quad \mbox{\cite{ABM:2013}}  \;, 
\label{ABM}
\end{eqnarray}
where the full NNLO QCD corrections evaluated in \cite{CM2012,CM2013}
have been combined with the soft-gluon resummation at NNLL accuracy \cite{Top++}
and Coulomb-gluon NNLL resummation  \cite{TOPIXS}.
All results in Eqs.~(\ref{m1})-(\ref{ABM}) have large theoretical uncertainties.
To improve the current precision of the top-mass determination from the total
cross section the higher order corrections as well as a reduction of PDF and $\alpha_s$
uncertainties are required \cite{pdf:1,pdf:2}.

\vspace{-0.5cm}
\section{Some additional sources of uncertainties}
\subsection{EW contribution to the running mass of the top-quark}
In order to achieve percent level precision theoretical predictions for 
cross section $\sigma_{pp \to t\bar{t}}$ not only QCD NNLO radiative
corrections should be applied. The EW part as well as mixed EW
$\times$ QCD corrections have to be included in a systematic way.  For
example, the QCD interaction is not responsible for the non-zero width
of the top-quark, which can be understood precisely only 
by inclusion of the EW interactions. 
In any case, the EW effect~\cite{Beenakker1993,Kuhn} 
as well as the non-zero width 
should be included in addition to the QCD corrections \cite{Signer}.

In contrast to QCD, where the mass of a quark is the parameter of the Lagrangian,  
the notion of \msbar-mass in EW theory is not determined entirely 
by the prescriptions of minimal subtraction. 
It depends on the value of vacuum expectation value $v(\mu^2)$ 
chosen as a parameter of the calculations
so that the running mass is $m_t(\mu^2) = 1/\sqrt{2} y_f(\mu^2) v(\mu^2)$. 
It has been shown in a series of papers \cite{JKV2001,JKV2002,JKV2003} 
that in the scheme with explicit inclusion of tadpoles  \cite{FJ1981}  
the RGE for the running vacuum expectation value $v_{MS}^2(\mu^2)$ 
which is defined as (see Section 4 in \cite{JK2003})
\begin{eqnarray}
v_{MS}^2(\mu^2) = \frac{1}{ \sqrt{2} G_F} \frac{1}{1-\overline{\Delta}r}
\Biggl[\frac{m_W^2 (\mu^2)}{M_W^2}\Biggr]
\Biggl[\frac{\alpha(M_Z)}{\alpha_{MS}(\mu^2)} \Biggr]
\Biggl[\frac{\sin^2 \theta^{MS}_W(\mu^2)}{\sin^2 \theta_W^{OS}}\Biggr]  \; ,
\end{eqnarray}
coincides with RGE for the classical definition of tree-level scalar potential vacuum
\begin{eqnarray}
\mu^2 \frac{d}{d \mu^2} v_{MS}^2(\mu^2) = 
\mu^2 \frac{d}{d \mu^2} \left[\frac{m^2(\mu^2)}{\lambda(\mu^2)} \right]\;, 
\end{eqnarray}
where $m^2$ and $\lambda$ are the parameters of the scalar potential, see Eq.~(\ref{potential}).
The asymptotic behavior and properties of this vacuum at low and high energies 
have been analysed in \cite{JKK2013,dis2013}. 
In particular, 
for a current values of Higgs and top-quark masses 
an IR point $\mu_{IR}$  close to the value of the Z-boson mass exist
such that 
$
\left. 
v_{MS}^2(\mu^2) 
\right|_{\mu \sim M_Z}  = (\sqrt{2} G_F)^{-1/2} = 246.22~\gv 
$
(see details in \cite{dis2013}). 
In the framework of the effective potential approach, 
a similar condition for the minimum of effective potential 
is imposed by hand (see details in \cite{Quiros:1995}).

\begin{figure}[h]
\centerline{ 
  \includegraphics[width=4.8cm]{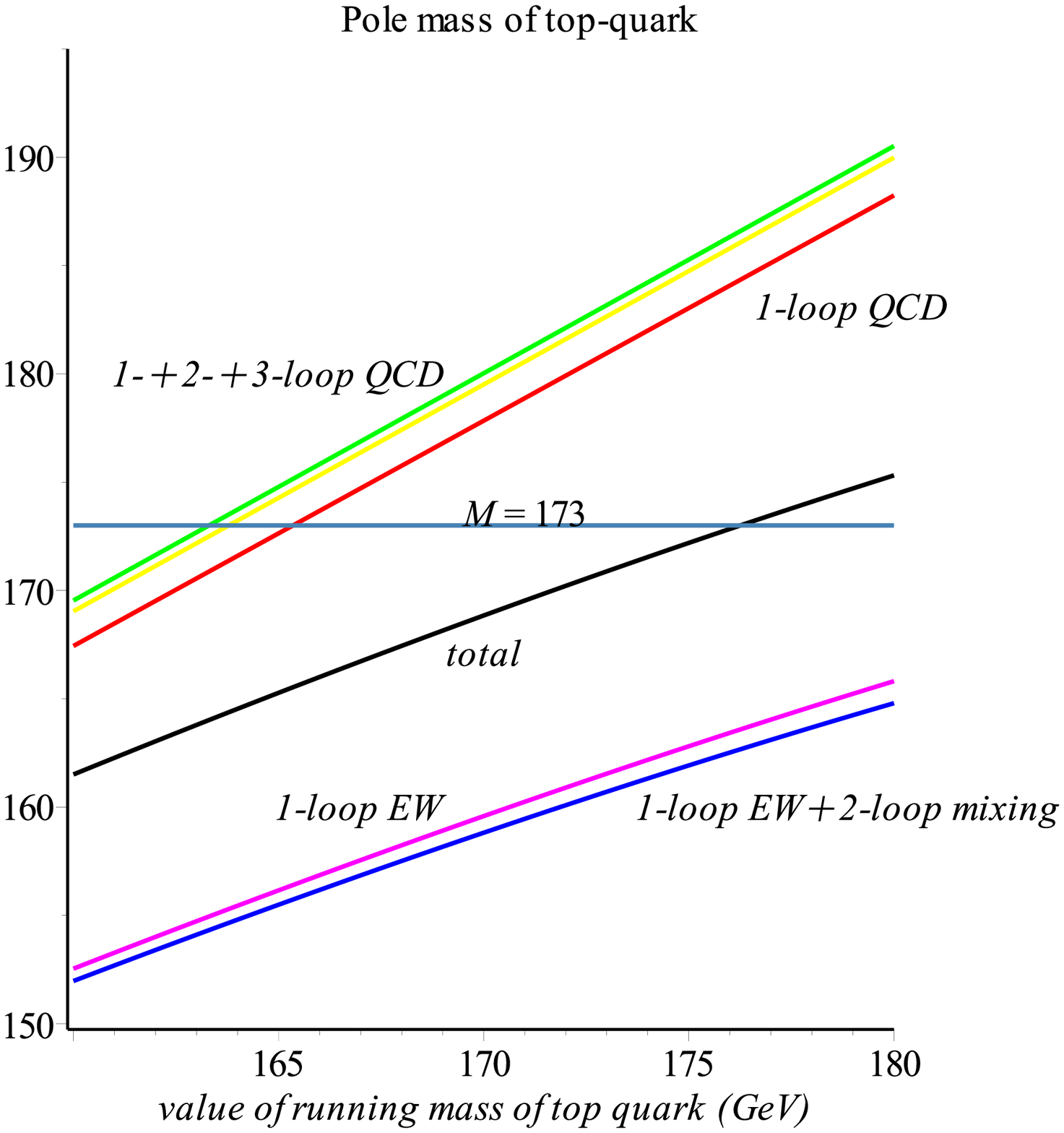}
  \hspace*{5mm}
  \includegraphics[width=4.8cm]{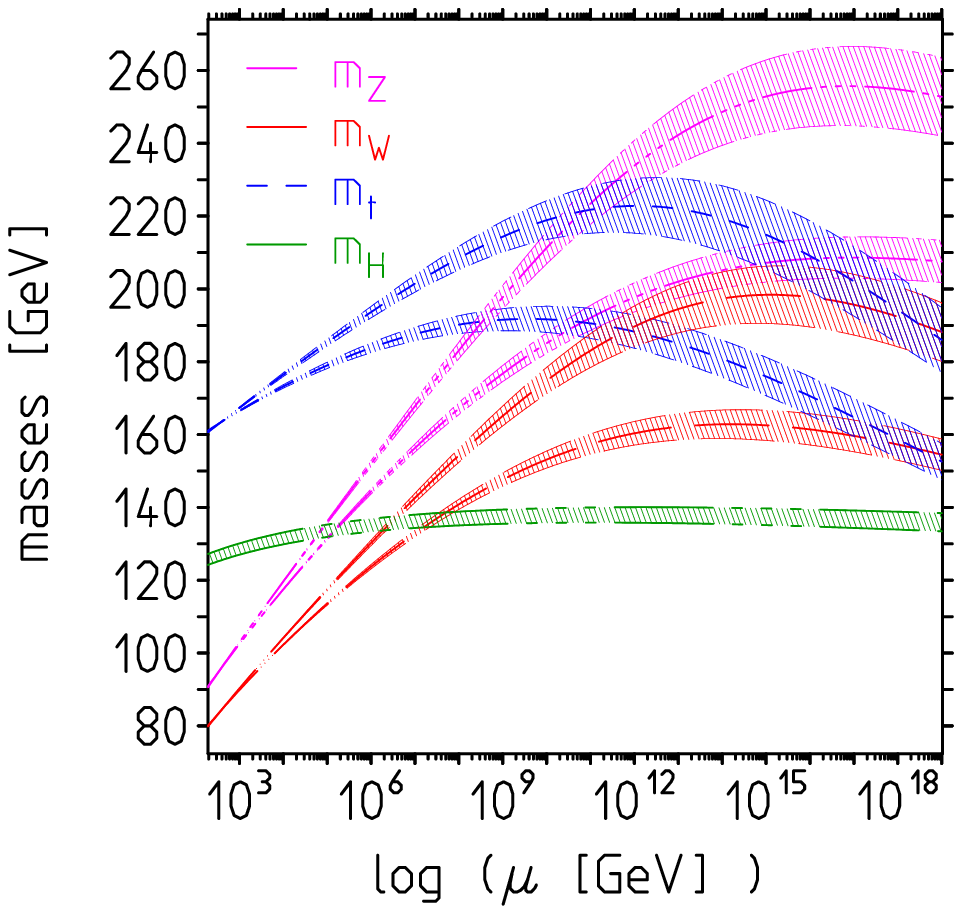}
            }
\vspace*{-3mm}
\caption{
The running top-quark mass at low value of $\mu$ (left)
and behavior of running masses at large values of $\mu$ (right).
At the right the bands corresponds to Higgs mass in interval
$124~\gv < M_H <127~\gv$.
The right plot is taken from \cite{Jegerlehner:2014}.}
\label{runningMass}
\end{figure}

In the SM the decoupling theorem~\cite{AC} is
not valid in the week sector, 
the ``decoupling by hand'' prescription does not work
and we have to take full SM parameter relations as they are.
In this approximation we got \cite{JKK2013} 
that the EW contribution is large and has opposite sign relative to the QCD 
contributions, so that the total SM correction is small and approximately 
equal to $M_t - m_t(m_t) \sim 1 \div 2 ~\gv$ (see left plot in Fig.~(\ref{runningMass})).
The complete $O(\alpha^2)$ correction to the relation between pole and
\msbar~~masses of the top-quark are not yet available, 
but our numerical estimation \cite{JKK2013} is 
in agreement with result of \cite{fermion,KV2014} and 
it is of the order of the $O(\alpha_s^4)$ corrections \cite{KK,Sumino}.
For the evaluation of the NNLO corrections three different schemes have been used:
with explicit inclusion of tadpole \cite{JKV2001,JKV2002,JKV2003}, 
the so called $\beta$-renormalization scheme \cite{Higgs3,Higgs3b,Higgs4},
and the method of minimization of the effective potential \cite{buttazzo2013,Degrassi:2012,Martin:2003}. 
The equivalence of these three methods of evaluation of matching conditions 
have never been analyzed. In particular, the structure of 2-loop UV-counterterms 
in  \cite{JKV2001,JKV2002,JKV2003} and \cite{Higgs3,Higgs3b} are different; 
the gauge dependence of the effective potential method versus  gauge independence of the scheme 
including the tadpoles  \cite{JKV2001,JKV2002,JKV2003} 
(the gauge (in)dependence of prediction of the critical value of the Higgs boson mass 
via effective potential scheme have been analysed in \cite{Willey:1997,Mihaila:2014}), etc. 
The difference between these methods is of order $O(G_F^2 M_W^4)$ 
and may reach the value $\sim 10^{-3} \div 10^{-4}$. 
Another source of numerical difference is related with the choice between
complex mass  \cite{Higgs4} or pole mass \cite{JKV2001,JKV2003} in Eq.~(\ref{mass}).
It is equivalent to systematic additional uncertainty of order $O(\Gamma^2/(4M^2))$ 
(see discussion in \cite{Smith,JK2004}), 
where $\Gamma$ is the width and $M$ is the mass of particle. 
This effect is the most significant for the top-quark Yukawa-coupling:
$\delta y_f \sim 1/173^2 =3 \times 10^{-5}$.
The unknown QCD corrections of order $O(\alpha_s^4)$ 
to the top-quark Yukawa-coupling is 
$\delta_{\alpha_s^4} y_f \sim 1 \times 10^{-3}$ \cite{KK,Sumino}.
The two-loop EW correction 
$\delta_{\alpha^2} y_t(\mu=M_t) \approx \frac{5}{(4 \pi)^2} = 6 \times 10^{-4}$ 
(see Eq.(2.49) in \cite{buttazzo2013})
and  three-loop  corrections can be estimated as
$\delta_{\alpha_s^2 \alpha} y_t(\mu=M_t) \sim 
[\delta_{\alpha_s^2} y_t(\mu=M_t)] \times [\delta_{\alpha} y_t(\mu=M_t)]
 = 0.011 \times 0.00135 = 1.5 \times 10^{-5}$.
All these effects do not change  the central value of the critical
mass of the Higgs boson, Eqs.~(\ref{2RG})-(\ref{full}), 
but affect the value of theoretical uncertainties. 

\vspace{-0.5cm}
\subsection{The first order phase transition and quadratic divergences} 
\label{1order}
It was pointed out in \cite{Jegerlehner:2014}
that in the region where 
the Higgs self-coupling is positive and close to zero, $\lambda(\Lambda) \sim 0$, 
the quadratic term of effective potential 
may start play the essential role since 
$\lambda(\Lambda) \phi^2(\Lambda)  \ll m^2(\Lambda).$
The massive  parameter  $m^2$ suffers from quadratic divergences:
$
\Delta m^2(\Lambda) = \frac{\Lambda^2}{16 \pi^2} Q_1 \;, 
$
where the coefficients $Q_1$ are expressible in terms of 
coupling constants \cite{Veltman:1981},
$
Q_1 = \lambda + \frac{1}{8} g_1^2 + \frac{3}{8} g_2^2 - y_t^2 \;. 
$
The coefficient $Q_1$ may vanish at some high scale $\Lambda^{*}$ 
\cite{Hamada:2012}.
If $\Lambda \sim \Lambda^{*}$ then $m^2$-term may change the sign 
and that leads to a phase transition of the first order 
restoring the EW symmetry \cite{Jegerlehner:2014}.
Realization of this scenario (the value of scale where $Q_1=0$)
strongly depends on the value of top-quark mass (see discussion in 
\cite{Masina:2013,Jones:2013}).
The Higgs inflation and hierarchy problem
within this scenario have been discussed in \cite{J:2014,J:2013}.  

\subsection{Inclusion of higher dimension operators} 
As well known \cite{Ford:1993} 
the effective potential contains also the higher dimensional operators (see Eq.~(\ref{rest})).
The impact of new physics interaction at Planck scale 
was analysed in \cite{Branchina:2013} by adding two higher dimensional
operators $\phi^6$ and $\phi^8$ suppressed by inverse powers of Planck scale $M_{Planck}$
to the Higgs potential.
It has been shown that higher dimension operators 
may change the lifetime of the metastable vacuum, $\tau$,
from $\tau = 1.49 \times 10^{714} T_U$ 
to   $\tau = 5.45 \times 10^{-212} T_U$, 
where $T_U$ is the age of the Universe.  

\section{Conclusion}
The main results of the LHC is the discovery of the Higgs boson.
The second important result is the absence of a signal of new physics. 
After the Higgs boson discovery the Standard Model is completed.
Since all parameters of the Standard Model are defined now experimentally,
one can analyze the extrapolation of the SM up to the Planck scale.
The results of the recent analysis can be summarized as follows: 
\begin{itemize}
\item 
The Standard Model is a self-consistent QFT 
that can be extrapolated from $M_W$ to $M_{Planck}$ 
since all SM couplings remain perturbative (no Landau pole)
in that range, see Fig.~\ref{running}.

\item
With the current precision in $M_H (\sim126~\gv)$ , 
the value of top-quark mass in accordance with CDF/DO/CMS/ATLAS
$M_t (\sim173~\gv)$ 
and 3-loop RG functions and 2-loop matching conditions, 
one concludes that the EW vacuum would most likely be {\it metastable} \cite{buttazzo2013}: 
the {\it stability condition} is 
$
M_H > (129.6 \pm 1.5)~\gv 
$
for a given value of top-quark mass.
In terms of top quark mass, the stability bound is \cite{Espinosa:2013}
$
M_t < (171.36 \pm 0.46) ~\gv. 
$

\item
It was shown in \cite{last-prediction}
that the lifetime of the electroweak vacuum is longer than the age of the Universe
for $M_H > 111~\gv$ so that \\
{\it Metastability of vacuum with very long lifetime 
     cannot be used as motivation for a New Physics} \\
A lot of effort has been made recently
to analyzed EW vacuum during inflation \cite{Kobakhidze:2013tn,Enqvist:2014bua}. 
The condition of vacuum (meta)stability imposes the constraint 
on the rate of inflationary  expansion that are in tension with BICEP2 result \cite{BICEP2}. 
However, the recent result from the Plank collaboration \cite{planck}
does not confirm the BICEP2 result \cite{BICEP2}
so that the further experimental verification is necessary.

%

\item
The result of the recent analysis \cite{ADM} reveals a large
theoretical uncertainties in the value of top-quark mass (see Eqs.~(\ref{m1})-(\ref{ABM})).
This leads to a large uncertainty $\sim 5~\gv$ in the critical value of the Higgs boson mass. 

\item
The mass of the Higgs boson is very close to the values of ``critical Higgs mass'', 
$M_{crit}$:
the ``multiple point principle'' \cite{FN:1995},
Higgs inflation \cite{Bezrukov:2007,Hamada:2014iga,Bezrukov:2014bra}, 
asymptotic safety scenario \cite{gravity}
The  explicit realization $M_H=M_{crit}$  
would be a strong indication for the absence of a new physics scale between the Fermi and Planck scales.

\item
To clarify the situation, 
more precise measurements of the coupling constants $\alpha_s, \lambda, y_t$ are needed. 
Unfortunately, there are no ways of directly measuring the Higgs self-coupling
and top-quark Yukawa coupling with high precision in the immediate future \cite{L1,L2,L3,L4}. 

\item
The higher order calculations are also desirable.
\end{itemize}

\noindent
Therefore, \\
{\bf precision determinations of parameters are more important 
than ever and a real challenge for experiments at the LHC and at a future ILC}.

\section{Acknowledgments}
We are thankful to S.Alekhin, A.Arbuzov, F.Bezrukov and S.Moch 
for the stimulating discussions. 
MYK would like to thank the organizers of the conference
ACAT2014 and specially to 
Andrei Kataev, Grigory Rubtsov
and 
Milos Lokajicek
for the invitation and for creating such a stimulating atmosphere.
The authors are indebted to anonymous referee for constructive and valuable comments.
This work was supported in part
by the German Federal Ministry for Education and Research BMBF through Grant No.\ 05~H12GUE,
by the German Research Foundation DFG through the Collaborative Research Center No.~676
{\it Particles, Strings and the Early Universe---The Structure of Matter and Space-Time}.

\section*{References}
\providecommand{\newblock}{}

\end{document}